\documentclass[12pt]{article}
\usepackage{amsmath,amssymb,graphicx,mathrsfs,hyperref,geometry}
\newcommand{\be}{\begin{equation}}
\newcommand{\ee}{\end{equation}}
\newcommand{\bea}{\begin{eqnarray}}
\newcommand{\eea}{\end{eqnarray}}

\def\({\left(} \def\){\right)}

\renewcommand{\baselinestretch}{1.2}
\begin{document}
\title{\vspace{-1.3in}
{A correspondence between \\ strings in the Hagedorn phase \\ and asymptotically de Sitter space}}

\author{\large Ram Brustein${}^{(1)}$, A.J.M. Medved${}^{(2,3)}$
\\
\vspace{-.5in} \hspace{-.5in}  \vbox{
\begin{flushleft}
$^{\textrm{\normalsize
(1)\ Department of Physics, Ben-Gurion University,
Beer-Sheva 84105, Israel}}$
$^{\textrm{\normalsize (2)\ Department of Physics \& Electronics, Rhodes University,
Grahamstown 6140, South Africa}}$
$^{\textrm{\normalsize (3)\ National Institute for Theoretical Physics (NITheP), Western Cape 7602,
South Africa}}$
\\ \small \hspace{1.07in}
ramyb@bgu.ac.il,\ j.medved@ru.ac.za
\end{flushleft}
}}
\date{}
\maketitle
\begin{abstract}

A correspondence between closed strings in their high-temperature Hagedorn phase and asymptotically de Sitter (dS) space is established. We identify a thermal, conformal field theory (CFT) whose partition function is, on the one hand, equal to the partition function of closed, interacting, fundamental strings in their Hagedorn phase yet is, on the other hand, also equal to the Hartle-Hawking (HH) wavefunction of an asymptotically dS Universe. The Lagrangian of the CFT is a functional of a single scalar field, the condensate of a thermal scalar, which is proportional to the entropy density of the strings. The correspondence has some aspects in common with the anti-de Sitter/CFT correspondence, as well as with some of its proposed analytic continuations to a dS/CFT correspondence, but it also has some important conceptual and technical differences. The equilibrium state of the CFT is one of maximal pressure and entropy, and it is at a temperature that is above but parametrically close to the Hagedorn temperature. The CFT is valid beyond the regime of semiclassical gravity and thus defines the initial quantum state of the dS Universe in a way that replaces and supersedes the HH wavefunction.  Two-point correlation functions of the CFT scalar field are used to calculate the spectra of the corresponding metric perturbations in the asymptotically dS Universe and, hence, cosmological observables in the post-inflationary epoch. Similarly, higher-point correlation functions in the CFT should lead to more complicated cosmological observables.
\end{abstract}
\newpage
\renewcommand{\baselinestretch}{1.5}\normalsize

\section{Introduction}

Because of the well-known correspondence between asymptotically anti-de Sitter (AdS) spacetimes and conformal field theories (CFTs) \cite{maldacena,GKP,WittenAdS,AdSRev}, along with the observation that the isometries of de Sitter (dS) space act as the conformal group on the dS boundary, it has long  been expected that a similar duality should exist between asymptotically dS cosmologies and a different class of CFTs \cite{strominger,dsCFTWitten,maldacenadS}.
This idea was first put forth by Strominger \cite{strominger} for the case of an eternal dS spacetime and then  later for that of an inflationary cosmology  \cite{strominger2,strominger3}. Since dS space has a spacelike asymptotic boundary, this framework  leads  to a timeless boundary theory and, consequently, a non-unitary CFT. One can perhaps view the boundary theory as a Euclidean CFT by considering certain analytic continuations of the standard AdS/CFT correspondence \cite{maldacenadS,HarloStanford}.

The detailed implementation of the dS/CFT correspondence began to take shape with a proposal by McFadden and Skenderis --- following from \cite{sken0} ---   that the CFT duals to domain walls in Euclidean AdS space  could be analytically continued into what would be the CFT duals to Lorentzian inflationary cosmologies \cite{sken1,sken2,sken3,sken4,sken5,sken6}. The first explicit realization of a dS/CFT duality from its AdS counterpart was presented by Hertog and Hartle in \cite{hert1} (and further developed by Hertog and others in, {\em e.g.},  \cite{hert2,hert3,hert4,hert5,hert6,hert7,hert8}), where the relation between AdS/CFT holography and the wavefunction of the inflationary Universe was made precise. The two approaches differ in that McFadden and Skenderis consider quantum fluctuations about a real classical geometry, whereas Hertog {\em et al.}  employ complex semiclassical saddlepoint solutions of the gravitaional and the matter path integral. In this sense, only the latter is proposing a quantum wavefunction of the Universe along the same lines as
that proposed by Maldacena \cite{maldacenadS}. On the other hand, both are similar in their treatments of cosmological perturbations and the late-time state.

In spite of these successful programs, some have argued that a direct relation between a dS/CFT correspondence and an explicit string theory relaization is  still lacking ({\em e.g.}, \cite{nogo}). For instance, there is significant evidence that ultraviolet completion of a stable dS space is incompatible with semiclassical quantum gravity \cite{vafadS,dvalidS,paltiswamp}. But, for a more optimistic viewpoint, as well as an update on recent progress,  see \cite{opti,maldastrings}.

The main purpose of the current paper is to make a  concrete proposal for a new type of dS/CFT correspondence;  one that is conceptually different than previous attempts.  Our proposed CFT dual is at finite temperature  and so is not obviously scale invariant, but we will nevertheless argue that it is.  The CFT is the theory of the so-called thermal scalar and, as an effective description of a multi-string partition function, has played an important role in understanding  the Hagedorn phase of string theory \cite{veep,bluff,AW,brandgas1,HP,DarVen,BarbRab}. The correspondence is substantiated by showing that,
when  the fields and parameters of the two theories are suitably matched, the partition function of the CFT is equal  to the Hartle--Hawking (HH) wavefunction \cite{HH}  of an asymptotically  dS Universe \footnote{ For a recent discussion of the HH wavefunction, see \cite{HHH}.}. This equivalence is established in the semiclassical regime for which the HH wavefunction can be defined.

We are interested in the case that the equilibrium state of the CFT is a thermal state of closed, interacting, fundamental  strings in their Hagedorn phase. Such a state of strings  is known to be one  possessing maximally allowed  pressure \cite {AW} and maximal entropy  \cite{saskatoon}. We have recently proposed that this state should describe the initial state of the Universe \cite{dSHigh}; the motivation being that a state of maximal entropy is just what is needed to resolve spacelike singularities such as the interior of an event horizon or the pre-inflationary Universe. The latter case leads to a duality connecting the string state to dS space and, as shown in our current discussion,  implies a duality between dS space and a thermal-scalar condensate.
\cite{BHfollies}. Indeed, previous studies by Silverstein and collaborators have discussed, in a very different context, how a tachyon condensate can be used to tame spacelike singularities \cite{silver1,silver2,silver3,silver4}.
Note though that this dS spacetime is the invented artifact of a late-time observer, who  wishes to explain  the state's origins and properties by imposing some form of semiclassical evolution. In our framework, this notion of dS space does not really exist, certainly not as a semiclassical state.

The equilibrium state is  maximally entropic  in the sense  that its  spatially uniform entropy density is equal to the square root of its  spatially constant energy density in Planck units and, thus, the former  density saturates the causal entropy bound \cite{ceb}.  On the dS side of the correspondence, maximal entropy translates into  the Gibbons--Hawking values of the entropy  within a cosmological horizon \cite{GHdS} and  the constant energy density is interpreted as a cosmological constant.  In previous articles, starting with \cite{inny}, we have interpreted the saturation of the causal entropy bound as indicating that such a state cannot be described by a semiclassical geometry. Nonetheless, the Lagrangian of the CFT can be used to calculate cosmological observables in spite of the lack of a semiclassical geometric description.  The Lagrangian that is presented  here extends a free energy that was first introduced in \cite{strungout,emerge} to describe Schwarzschild black hole (BH) interiors. This free energy is expressed  as a power series in the entropy density and has a form that was adapted from the free energy of polymers ({\em e.g.}, \cite{polytext,pt2,pt3}).

Having identified the CFT dual for dS space, we can calculate correlations functions in the CFT  and then translate these into cosmological observables in the post-inflationary epoch without relying on semiclassical dS calculations. Our focus is on calculating the power spectra for  the tensor and scalar perturbations. We  have already presented qualitative expressions for these scale-invariant spectra in \cite{dSHigh}, but  the  CFT improves on this by  providing  a precise prescription for the relevant calculations. The results presented here are shown to be in agreement with those of  standard inflationary calculations \cite{mukhanov} and with those obtained using the HH wavefunction \cite{Halliwell:1984eu,Hartle:2010vi,Hertog:2013mra}.

Our proposed model has some features in common with those of string-gas cosmology as presented by Brandenberger, Vafa and collaborators \cite{brandgas1,brandgas2,brandgas3,brandgas4,brandgas5}, as well as with the  holographic cosmology  model of Banks and Fischler \cite{BF1,BF2}. However, as discussed at length in \cite{dSHigh}, such similarities are mostly superficial due to differneces both in the physical substance and in the resulting predictions.

A main difference between our proposal for a dS/CFT duality  and previous ones is that ours does not rely upon an intermediating semiclassical Euclidean AdS solution. Also, our  duality is holographic but not in the usual way: It is holographic in the sense that the thermal scalar condenses by winding around a compactified Euclidean time loop. So the thermal scalar field theory ``lives" in one less dimension. This is not equivalent to taking a small limit $S^1$ in a bulk gravity description. Strings are essential, as there is no condensate as a matter of principle without strings winding on a string-length-sized thermal circle. It is also worth emphasizing that there is a clear string-theory origin for our model because, as is well known, the effective field theory of the thermal scalar can be used to calculate the string partition function near the Hagedorn temperature.

Briefly on the contents, the next section introduces the CFT Lagrangian, Section~3  discusses the various aspects of the theory in terms of thermal-scalar condensate and  Section~4 establishes the correspondence to dS space. We then present our calculations of the cosmological observables in Section~5 and conclude in Section~6.

\section{Thermal scalar of closed strings in the Hagedorn phase}

Let us begin here with the quantum partition function
for closed, interacting strings  $\; Z=Tr e^{-\beta H}\;$, where  $H$ is the Hamiltonian and $\beta$ is related to the temperature $T$ as in Eq.~(\ref{tempdef}).   The partition function and  its associated thermal expectation values  can be calculated  in terms of a Euclidean action $\mathcal{S}_E$ that is obtained by  compactifying imaginary time on a ``thermal circle'',
\be
\mathcal{S}_E\;=\;\oint_0^{\beta} d\tau\; \sqrt{g_{\tau\tau}}\int d^{d}x \; \sqrt{\gamma}~ \mathcal{L}_E\;,
  \ee
where
\be
\;\frac{1}{T}\;=\;\oint_0^{\beta}d\tau\;\sqrt{g_{\tau\tau}}\;,
\label{tempdef}
\ee
and where the $\;D=d+1$-dimensional coordinate system and metric tensor should be regarded as those of a  fiducial manifold, since the string state lacks a semiclassical geometry.
We will be  discussing the case in which  temperatures are close to but slightly above the Hagedorn temperature, $\;T\gtrsim T_{Hag}\;$ and  $\;T-T_{Hag} \ll T_{Hag}\;$.
It follows that the circumference of the thermal circle is on the order of the string length $l_s$.

Compactifying time and ignoring the time-dependence of the fields  amounts to reducing the dimensionality of the theory from $d+1\;$ to $d$. The result is then a ``timeless'' theory living on a $d$-dimensional spatial hypersurface, just as expected from a would-be dS/CFT correspondence.

Strings can wind around the thermal circle and the resulting picture can be described by using the  well-studied theory of the thermal scalar  \cite{veep,bluff,AW,brandgas1,HP,DarVen,BarbRab}.  The $+1$ winding mode is denoted by $\phi$ and its $-1$ counterpart is denoted by $\phi^*$.  As the winding charge is a conserved quantity, the Lagrangian is required  to be a functional of $|\phi|^2$. The path integral of the thermal scalar is known to provide an effective (but complete) description of the multi-string partition function  when the temperature is close to the Hagedorn temperature.

The   Lagrangian  of the thermal scalar can be expressed as
\be
{\cal L}_E(\phi,\phi^*)\;= \;{ \textstyle\frac{1}{2}}  \gamma^{ij}\partial_i \phi \partial_j \phi^* -c_1 \; \varepsilon\; T \phi\phi^*+{\textstyle \frac{1}{2}} c_2\; g_s^2\; T^2 \left(\phi\phi^*\right)^2+\cdots\;,
\label{LE2}
\ee
where $\;\varepsilon=T-T_{Hag}\;$, $g_s^2$ is the dimensional string-coupling constant and the positive, dimensionless numerical coefficients $c_1$ and $c_2$ depend on the specific string theory. The ellipsis denotes higher-order interactions, both here and below (and will sometimes be omitted). The relative unimportance of these higher-order terms will be discussed in the next section.  The potential for the thermal scalar was introduced a long time ago in \cite{AW}. We
 have made  here a choice of sign that ensures a non-trivial solution in the regime of interest (see below). The total mass dimension of the Lagrangian density has to, of course, be $d+1$. Because the mass dimension of $\varepsilon$ is $+1$ and that of the dimensional coupling  $g_s^2$ is $-(d-1)$, it  then follows that the mass dimension of $\phi$ is $+\frac{d-1}{2}$. We may absorb the numerical coefficients  by the redefinitions $\;c_1\varepsilon \to  \varepsilon\;$ and $\; c_2 g_s^2 \to  g_s^2\;$, thus giving
\be
{\cal L}_E(\phi,\phi^*)\;= \;{ \textstyle\frac{1}{2}}  \gamma^{ij}\partial_i \phi \partial_j \phi^* - \varepsilon\; T \phi\phi^*+{\textstyle \frac{1}{2}}  g_s^2\; T^2 \left(\phi\phi^*\right)^2\;.
\label{LE1}
\ee

For temperatures below the Hagedorn temperature ($\varepsilon < 0$), the thermal scalar is known to have a positive mass-squared \cite{AW}. Meanwhile, its mass vanishes at Hagedorn transition temperature $\varepsilon=0$, and so it is tempting to adopt the standard viewpoint that the phase transition is describing the condensation of closed-string winding modes about the thermal circle. This perspective is especially interesting for the case of BHs,
as it aligns nicely with earlier proposals that a Euclidean BH --- albeit one in an AdS spacetime --- could be related to the condensation of the thermal scalar \cite{BR1,BarbRab,BR3,ofer,notBR1,notBR2,SunnyNew}. However, as should become clear by the end of the section, the Lagrangian~(\ref{LE1}) has to be regarded as an expansion near a non-trivial minimum of the potential which lies {\em above} the
Hagedorn temperature.  The restriction to  trans-Hagedorn temperatures can
understood by noticing that the entropy and energy densities both vanish for $\;\varepsilon=0\;$ ({\em cf}, Eqs.~(\ref{consol}-\ref{rhostring})) and that the former density formally becomes negative for $\;\varepsilon<0\;$. Hence, the Lagrangian~(\ref{LE1})  cannot be used directly to describe the Hagedorn phase transition and reproduce its expected first-order character.

The equation of motion $\;\phi^*\delta {\cal L}_E/\delta\phi^*=0\;$ is as follows:
\be
-\frac{1}{2} \phi^*\nabla^2\phi -\varepsilon\; T \phi\phi^*+ g_s^2\; T^2 \left(\phi\phi^*\right)^2 \;=\;0\;.
\ee
An interesting solution of the above equation and its conjugate is one in which the thermal scalar condenses,
\be
|\phi_0|^2\;=\;\frac{\varepsilon}{g_s^2\; T}\;.
\ee
It will be shown later that this ratio is a small number in comparison to the Hagedorn scale, $\;\varepsilon/(g_s^2\; T_{Hag} ) \ll T_{Hag}^{d-1}\;$.

Expanding the Lagrangian about this constant solution, $\;\phi=\phi_0+\varphi\;$, $\;\phi^*=\phi_0+\varphi^*\;$, we find that
\be
{\cal L}_E
\;=\;{\textstyle \frac{1}{2}} \gamma^{ij}\partial_i \varphi \partial_j \varphi^* + \varepsilon T \varphi\varphi^*+{\textstyle \frac{1}{2}}  g_s^2 T^2 \left(\varphi\varphi^*\right)^2 -\frac{1}{2} \frac{\varepsilon^2}{g_s^2}\;.
\ee
One may also include  a coupling to the Ricci scalar in the Lagrangian. For instance, if a conformal coupling is chosen, then   $\;{\cal L}_E\to {\cal L}_E-{\textstyle \frac{d-1}{4d}} R \varphi \varphi^*\;$.  The importance of this inclusion will be revealed later on; however,  as one always has the freedom to choose Ricci-flat fiducial coordinates, this term cannot be relevant to the calculation of physical observables.

The expanded Euclidean action is thus given by
\be
\mathcal{S}_E\;=\; \frac{1}{T}\int d^d x \sqrt{\gamma} \left\{
{\textstyle \frac{1}{2}} \gamma^{ij}\partial_i \varphi \partial_j \varphi^*-{\textstyle \frac{d-1}{4d}} R \varphi \varphi^* +\varepsilon\; T \varphi\varphi^*+{\textstyle \frac{1}{2}}  g_s^2\; T^2 \left(\varphi\varphi^*\right)^2 -\frac{1}{2} \frac{\varepsilon^2}{g_s^2} \right\}\;.
\label{finallagphi}
\ee
The action  in Eq.~(\ref{finallagphi}) is  similar to the standard expression in the literature ({\em e.g.}, \cite{AW,HP}).

\section{Thermal scalar condensate}

In this section, we elaborate on  some of the consequences for our theory when the thermal scalar condenses.

\subsection{Euclidean action}
In the case of condensation, it is simpler to use the real field
\be
s\;=\;|\phi|^2\; T
\ee
as the fundamental field; for which the  expectation value at the minimum is then
\be
s_0 \;=\; \frac{\varepsilon}{g_s^2}\;.
\ee
We have denoted the  field  by $s$   because its condensate value $s_0$ is the same   as the local entropy density of the strings (see below).

Let us now rewrite the  Lagrangian~(\ref{LE1}) as a functional of $s$,
\be
\mathcal{L}_E(s)\;=\; \frac{1}{8}  \;\frac{1}{s T} \gamma^{ij}\partial_i s\partial_j s  -\varepsilon s +{\textstyle \frac{1}{2}}g_s^2 s^2\;.
\label{lag1}
\ee
Expanding  the above  near the minimum $\;s= s_0\left(1 + \sigma(x_i)\right)\;$, keeping only quadratic terms and recasting it as a compactified Euclidean action  as in Eq.~(\ref{finallagphi}), we  have
\be
\mathcal{S}_E^{(2)} \;=\; \frac{1}{T} \int d^d x \sqrt{\gamma} {\cal L}_E (\sigma)  + \mathcal{S}_0 \;,
\ee
such that
\be
\mathcal{S}_0 \;=\; - \frac{1}{T}\int d^d x\sqrt{\gamma}~ \frac{1}{2} \frac{\varepsilon^2}{g_s^2}
\label{sos}
\ee
and
\be
\mathcal{S}_E^{(2)} \;=\;\frac{1}{g_s^2 T} \int d^d x \sqrt{\gamma}\bigg\{{\frac{1}{8}} \frac{\varepsilon}{T} \gamma^{ij}\partial_i \sigma\partial_j \sigma + \frac{1}{2} \varepsilon^2\sigma^2 - \frac{d-1}{16 d} \frac{\varepsilon}{T}  R \sigma^2
\bigg\}+ \mathcal{S}_0\;,
\label{lag22}
\ee
with the conformal coupling to $R$ included for completeness.

The equation of motion that results  from the action~(\ref{lag22}), for the case of Ricci flatness, is found to be
\be
-\nabla^2 \sigma +4\; \varepsilon\; T \sigma \;=\; 0\;.
\label{eom}
\ee
The field $\sigma$ is therefore  a massive, conformally coupled scalar with a positive thermal mass-squared, $\;m^2=+4\; \varepsilon\;T$. This value for $m^2$
can  be compared with the magnitude of the negative mass-squared of the thermal scalar when it is below the Hagedorn temperature, $\;m^2=-\varepsilon\; T\;$ ({\em e.g.}, \cite{HP}).

We may absorb the dimensionality of $g_s^2$ and $\varepsilon$ by rescaling them with appropriate powers of the temperature,
\be
\widetilde{g}_s^{\; 2} \;= \;g_s^2 T^{d-1}\;,
\ee
\be
\epsilon \;=\; \frac{\varepsilon}{T}\;.
\ee
In which case,
\be
\mathcal{S}_E^{(2)} \;=\;\frac{1}{\widetilde{g}_s^{\; 2} } T^d \int d^d x \sqrt{\gamma}\bigg\{{\frac{1}{8}}\; \epsilon\; \frac{1}{T^2}\gamma^{ij}\partial_i \sigma\partial_j \sigma + 2 \epsilon^2\sigma^2 - \frac{d-1}{16 d}\;\epsilon\; \frac{1}{T^2}  R \sigma^2
\bigg\}\;.
\label{action2fin}
\ee
As the field $\sigma$ is dimensionless by its definition,  the only remaining dimensional parameter is $T$, making this a thermal CFT. We will explain  how scale and Weyl transformations act on this action, after   discussing the higher-order interactions.

Higher-order (HO) terms in the action come about in two different ways: (I) more than two strings intersecting at a single point  or (II) the same pair of  strings intersecting at two or more different points.
Additional action terms of the former kind are
\be
\mathcal{S}_E^{(HO,I)} =\frac{1}{T} \int d^d{x} \sqrt{\gamma}\bigg\{{\frac{a_3}{3!}}\frac{1}{T}(g_s^2)^2 s^3 + {\frac{a_4}{4!}}\frac{1}{T^2}(g_s^2)^3 s^4  + \cdots\bigg\}\;,
\label{SHO34}
\ee
where the $a$'s (and $b$'s below) are numerical coefficients and the additional powers of  temperature are dictated by the scaling dimensions of the various quantities.
Expanding about the minimum $\;s= s_0(1 + \sigma(x_i))\;$, we then have
\bea
\mathcal{S}_E^{(HO,I)} &=&\frac{1}{{g}_s^2 T} \int d^d{x} \sqrt{\gamma}\bigg\{{\frac{a_3}{3!}}\varepsilon^2 {\epsilon}(1+\sigma)^3 + {\frac{a_4}{4!}}\varepsilon^2 \epsilon^2(1+\sigma)^4  + \cdots\bigg\}
\label{SHO34b} \nonumber \\
&=&\frac{T^d}{\widetilde{g}_s^{\; 2} } \int d^d x \sqrt{\gamma}\bigg\{{\frac{a_3}{3!}}{\epsilon^3}(1+\sigma)^3 + {\frac{a_4}{4!}} \epsilon^4(1+\sigma)^4  + \cdots\bigg\}\;,
\label{HOI}
\eea
where all parameters and fields besides $T$ are explicitly  dimensionless in the lower line.  As the small expansion parameter in this case is $\;\epsilon=
\frac {T-T_{Hag}}{T} \ll 1\;$, these corrections can
be identified as $\alpha^\prime$ corrections in the effective action.

Higher-order terms coming from the same strings intersecting at two or more different  points take the form
\bea
\mathcal{S}_E^{(HO,II)} \;=\;\frac{1}{T} \int d^d{x} \sqrt{\gamma}\bigg\{ \frac{b_2}{2!} T^{d-1}(g_s^2)^2 s^2 + \frac{b_3}{2!} T^{2(d-1)}(g_s^2)^3 s^2   + \cdots\bigg\}\;.
\label{SHOg1}
\eea
Once again expanding about the minimum and converting to dimensionless quantities, we obtain
\bea
\mathcal{S}_E^{(HO,II)} &=&\frac{1}{T g_s^2} \int d^d{x} \sqrt{\gamma}\bigg\{ \frac{b_{2}}{2!} T^{d-1}(g_s^2) \varepsilon^2 (1+\sigma)^2 + \frac{b_3}{2!} T^{2(d-1)} (g_s^2)^2 (1+\sigma)^2   + \cdots\bigg\}
\label{SHOg2} \nonumber  \\
&=&\frac{T^d}{\widetilde{g}_s^{\; 2}} \int d^d x \sqrt{\gamma}\bigg\{ \frac{b_{2}}{2!} \widetilde{g}_s^{\; 2} \epsilon^2 (1+\sigma)^2 + \frac{b_3}{2!}(\widetilde{g}_s^{\; 2})^2 \epsilon^2 (1+\sigma)^2   + \cdots\bigg\}\;.
\label{SHOgfin}
\eea
The small expansion parameter in this case is $\widetilde{g}_s^{\; 2}=g_s^2 T^{d-1}$, and so these  are identifiable as string loop corrections in the effective action.

There are, of course, more complicated higher-order interaction terms involving both string-coupling and $\alpha^\prime$ corrections. All of these corrections are parametrically small provided that the requisite hierarchy $\;\epsilon \ll \widetilde{ g}_s^{\: 2} < 1\;$  (see Subsection~3.3)
is respected.

\subsection{Conformal symmetry}

Let us now discuss the transformation properties of the theory under Weyl transformations. We first restrict attention to the case of constant Weyl transformations, which  correspond to scale transformations of the coordinates. For the $d+1$-dimensional Euclidean theory,  the constant Weyl transformations can be expressed as
\bea
g_{\tau\tau}\; &\to\; & \Omega^{2} g_{\tau\tau}\;, \cr
\gamma_{ij}\;&\to\;& \Omega^{2} g_{ij}\; .
\eea

As we have seen,
the  dimensional coupling parameters $g_s^2$ and $\varepsilon$ can be rendered dimensionless by rescaling
them with appropriate powers of the temperature, as done in Eqs.~(\ref{action2fin}),~(\ref{HOI}) and~(\ref{SHOgfin}). Meaning  that the only remaining dimensional parameter is the temperature. The question then is how  to interpret the parameter $T$ in the $d$-dimensional compactified theory.   If one considers the temperature to  be a fixed dimensional parameter, then this is obviously not  a scale-invariant theory. However, if one rather considers that the temperature  is the inverse of the circumference of the thermal circle as in Eq.~(\ref{tempdef}), $\;\frac{1}{T}=\oint_0^{\beta}d\tau\;\sqrt{g_{\tau\tau}}$\;, then it obviously varies under a Weyl transformation as
\be
T \;\to\; T/\Omega\;.
\label{Ttrans}
\ee
Then, in this case, the variation of the metric in each of Eqs.~(\ref{action2fin}),~(\ref{HOI}) and~(\ref{SHOgfin}) is exactly canceled by the variation of the temperature, as the product $T^d\sqrt{\gamma}$,  in particular, is scale invariant. Since the zeroth-order part of the action in Eq.~(\ref{sos}),
$\;\mathcal{S}_0 = {T^d}\int d^d x\sqrt{\gamma}\frac{1}{2} \frac{\epsilon^2}{{\widetilde g}_s^2}\;$, transforms similarly, the complete action is scale invariant.

When the temperature varies as in Eq.~(\ref{Ttrans}), the theory is also invariant under general $x$-dependent Weyl transformations,
\bea
g_{\tau\tau}\; &\to\; & \Omega^{2}(x_i) g_{\tau\tau}\;, \cr
\gamma_{ij} &\;\to\; & \Omega^2(x_i) \gamma_{ij}\;.
\eea
The only term that is sensitive to the difference between constant and $x$-dependent Weyl transformations is the kinetic term. However, the conformal coupling of the scalar to the Ricci scalar ensures the invariance of the kinetic term  even under spatially dependent Weyl transformations.  It can then be concluded that, when the parameter $T$ varies according to Eq.~(\ref{Ttrans}), the thermal-scalar condensate is described by a CFT, in spite of the appearance  of a  dimensional scale --- the temperature.

\subsection{Free energy and thermodynamics}

For the physical interpretation of the condensate solution, it is helpful to recall our previous discussions on the Helmholtz free energy of strings that are slightly above the Hagedorn temperature \cite{strungout,emerge}. There, we proposed a free energy density  which  is similar to those of polymers with attractive interactions ({\em e.g.}, \cite{polytext,pt2,pt3}). In particular, the free energy density $F/V$ should be regarded as an expansion
in terms of the entropy density $s$ such that  $\;s\ll T^d_{Hag}\;$,
\be
-\left(\frac{F}{V}\right)_{strings}\;= \;\varepsilon s - \frac{1}{2} g_s^2   s^2 + \cdots\;,
\label{FES2}
\ee
where the ellipsis, as usual, denotes higher-order interaction terms.
The right-hand side of Eq.~(\ref{FES2})  is the same as  the potential in Eq.~(\ref{lag1}).

From this stringy point of view,  $\varepsilon$  should  be regarded as the strings' effective temperature. That is, the temperature associated with the collective motion of long strings, rather than the  local value of the temperature of small pieces of string (or ``string bits") for which the temperature is much higher, $\;\varepsilon \ll T\sim T_{Hag}\;$.

The first term on the right of Eq.~(\ref{FES2}) represents the Helmholtz free energy  of a  free string. In the free case and in string units ($l_s=1$), both the energy $E$ and the entropy $S$ are equal to the total length $L$ of the strings, $\;E=L\;$ and $\;S=L\;$. It follows that $\;F/V= (E-ST)/V=(1-T)L/V\;$ and then, since $\;s=S/V=L/V\;$ and $\;\varepsilon=T-T_{Hag}\;$, also that $\;F/V \simeq -\varepsilon s \;$, where  we have approximated $\;T\simeq T_{Hag}\simeq 1/l_s=1\;$.

The second term on the right of Eq.~(\ref{FES2}) --- the leading-order interaction term ---  can be understood by recalling that a closed string interacts at its intersections, either with itself or with another string. The simplest such interactions being those for which two closed strings join to form one longer
one or one closed string splits into two shorter ones. Since the  probability of interacting  is given by the dimensionless string-coupling constant $\widetilde{g}_s^{\; 2}$,  and  again under the assumptions that  $\;T\sim T_{Hag}\sim 1\;$ and that any numerical or phase-space factors were absorbed into the dimensional coupling, the total interaction strength is proportional to $\;\widetilde{g}_s^{\; 2} L^2/V = \widetilde{g}_s^{\; 2} s^2 V\;$.   As for the higher-order terms, these will include  extra factors  of  $\;\widetilde{g}_s^{\; 2} L/V\sim \widetilde{g}_s^{\; 2} s\sim \epsilon\;$ (see Eq.~(\ref{consol}) below) and/or $\widetilde{g}_s^{\; 2}$ when the same strings intersect at multiple points. Therefore,  $\;\epsilon,\;\widetilde{g}_s^{\; 2} < 1\;$ are  necessary conditions for these interactions to be suppressed. Equation~(\ref{consol}) below further implies the hierarchy  $\;\epsilon\ll \widetilde{g}_s^{\; 2} <1\;$.

The minimization of the free energy defines the equilibrium state. Doing so, one obtains
what was previously identified as the condensate solution,
\be
\label{consol}
s \;=\; \frac{\varepsilon}{g^2_s}\;,
\ee
which along with standard thermodynamics (with $\varepsilon$ serving as the temperature) yields the equilibrium relations
\be
\label{rhostring}
p\;=\;\rho \;=\;  \frac{1}{2}\frac{\varepsilon^2}{g^2_s}\;,
\ee
where  the first equality is independent of Eq.~(\ref{consol}).  The causal entropy bound is indeed parametrically saturated since $\;s\sim\sqrt{\rho}\;$.

\subsection{An effective two-dimensional conformal field theory}

As previously discussed, the thermal-scalar condensate can be viewed as a $d$-dimensional Euclidean CFT. However, as we  now show, it is effectively a two-dimensional CFT. This aspect of the thermal scalar  was noticed a long time ago in \cite{saskatoon} and is implicit in \cite{AW}. We have already discussed this feature of the theory in the context of BHs in \cite{strungout,emerge}.

The free energy density of a $D$-dimensional (Euclidean) CFT  at temperature $1/\beta$ is expressible as~\footnote{In this subsection, we often adopt  notation from \cite{1802}.}
$\;
F/V= f_{\beta} \beta^{-D}\;,
$
where $f_{\beta}$ is a numerical coefficient. This  leads to an energy density of the form
$\;
\rho  =  - \left(1-\frac{1}{D}\right) b_\beta \beta^{-D}\;,
$
with $b_{\beta}$ being another number.
The two  coefficients are related according to  $\;f_{\beta}= {b_\beta}/{D}\;$ and an expression for the  entropy density $s$ promptly follows, $\;s=-b_{\beta}\beta^{-(D-1)}\;$.

For the case of $\;D=2\;$,
\bea
\label{cft2d1}
F_2/V &=& \tfrac{1}{2} (b_\beta)_2 \beta^{-2}\;, \\
\label{cft2d2}
\rho_{2} &=& -\tfrac{1}{2} (b_\beta)_2\; \beta^{-2}\;, \\
s_2 &=& -(b_\beta)_2\; \beta^{-1}\;.
\label{cft2d3}
\eea

Whereas, in our case,
\bea
\label{therm2d1}
F/V &=& -\frac{1}{2} \frac{\varepsilon^2}{g_s^2}\;, \\
\label{therm2d2}
\rho &=& \frac{1}{2} \frac{\varepsilon^2}{g_s^2}\;,\\
s &=& \frac{\varepsilon}{g_s^2}\;.
\label{therm2d3}
\eea

 Identifying $\varepsilon$ as the effective temperature,
\be
\varepsilon\;=\; 1/\beta\;,
\label{epsbeta}
\ee
and setting
\be
(b_\beta)_2 \;=\; - 1/g_s^2\;,
\ee
one can see a perfect match between  Eqs.~(\ref{therm2d1})-(\ref{therm2d3}) and Eqs.~(\ref{cft2d1})-(\ref{cft2d3}).

Moreover, if we adopt the  standard parametrization for the energy density of  a two-dimensional CFT in terms of the central charge  $c$,  $\;\rho = \frac{\pi}{6} c \beta^{-2}\;$ (see, {\em e.g.}, \cite{pap11}), then
\be
c \;=\;\frac{3}{\pi} \frac{1}{g_s^2}\;,
\label{center1}
\ee
and  it follows that
\be
s\;=\; \frac{\pi}{3} c \beta^{-1}\;.
\ee
The relations  $\;c\sim 1/g^2\;$  and $\;s\sim c\;$ are indeed universal features of CFTs, whereas the numerical coefficients depend on additional detailed information. The central charge is also expected to be related to the two-point function of the stress--energy tensor as
$\;\langle T^0_{~0} T^{0}_{~0}\rangle\sim c\;$.  This will be verified in detail next.

In CFTs at finite temperature, an operator with a non-vanishing conformal dimension can have a non-zero expectation value ({\em i.e.}, a thermal one-point function),
\be
\langle O \rangle_\beta = \frac{A_O} {\beta^{\Delta_O}}\;,
\ee
where  $\Delta_O$ is the conformal dimension and $A_O$ is a dimensionless coefficient for the operator $O$.
The scaling of such a one-point function can be specified  in terms of the stress--energy tensor,
\be
\frac {\partial \langle O \rangle_\beta}{\partial \beta}\;=\; -\frac{1}{\beta} \int d^{d+1} x \langle T^0_{~0}(\vec{x}) O(0)\rangle_{\beta}^c\;,
\label{confD}
\ee
where the superscript $c$ signifies a connected  function.

Choosing $O$ as the stress--energy tensor itself, one obtains
\bea
\frac{\partial \langle T^0_{~0} \rangle_\beta}{\partial \beta}&=& -\frac{1}{\beta} \int d^{d+1} x \langle T^0_{~0}(\vec{x}) T^0_{~0}(0)\rangle_{\beta}^c \nonumber \\
&=& - \int d^{d} x \langle T^0_{~0}(\vec{x}) T^0_{~0}(0)\rangle_{\beta}^c \;,
\label{confD2}
\eea
where the time circle has  now been compactified to a circumference of
  $\;\beta=1/\varepsilon\;$  so as to agree with the  definition of
the stress--energy tensor.
Both sides of Eq.~(\ref{confD2}) have explicit expressions in the CFT, and  so we can verify the relationship directly, a highly unusual situation for interacting CFTs.

First, using  Eqs.~(\ref{therm2d2}),~(\ref{epsbeta}) and the Euclidean identification
$\;T^0_{~0}=\rho\;$, one can translate the left-hand side of Eq.~(\ref{confD2})
into
\be
\frac{\partial \langle T^0_{~0} \rangle_\beta}{\partial \beta}\; = \;- \frac{\varepsilon^3}{g_s^2}\;.
\label{lefty}
\ee

The evaluation  of the right-hand side of Eq.~(\ref{confD2}) requires some additional ingredients.
Since the Euclidean action is expressed in terms of the entropy density $s$, a direct relationship between $T^0_{~0}$ and $s$ is required. For this, recalling that $\varepsilon$ is the effective temperature, we rely on the thermodynamic relation $\;\delta \rho = \varepsilon\;\delta s\;$. It follows that
\be
T^0_{~0}(\vec{x})- \langle T^0_{~0}(\vec{x}) \rangle_\beta\; =\; \varepsilon \left(s(\vec{x})- \langle s(\vec{x}) \rangle_\beta\right)\;,
\ee
and so
\be
\langle T^0_{~0}(\vec{x}) T^0_{~0}(0)\rangle_{\beta}^c \;=\;
\varepsilon^2 \langle s(\vec{x}) s(0)\rangle_{\beta}^c\;.
\label{t002}
\ee

We are interested in the limit $\;|\vec{x}| \varepsilon \gg 1\;$, as this will later be shown to describe super-horizon scales. In this case, the Euclidean action reduces to a single term, as can be seen from Eq.~(\ref{lag22}),
\be
S_{E} \;\sim\; \beta\int d^d x\; \tfrac{1}{2} g_s^2 (s - \langle s\rangle)^2\;.
\label{gauss1}
\ee
The two-point function of $s$ can then be readily  evaluated in terms of a Gaussian integral, again using $T=\varepsilon$,
\be
\langle s(\vec{x}) s (0) \rangle_{\beta}^c \;=\;\int {\cal D}[s]\; s(\vec{x}) s (0) e^{-\mathcal{S}_E(s;\beta)}\; \;=\; \frac{\varepsilon}{g_s^2}\delta^d(\vec{x})\;,
\label{gauss2}
\ee
which, by way of Eq.~(\ref{t002}), leads to
\be
\langle T^0_{~0}(\vec{x}) T^0_{~0}(0)\rangle_{\beta}^c \;=\;
 \frac{\varepsilon^3}{g_s^2}  \delta^d(\vec{x})\;.
\label{confD4}
\ee

It can now be  verified that the right-hand side of Eq.~(\ref{confD2}),
\be
- \int d^d x \langle T^0_{~0}(\vec{x}) T^0_{~0}(0\rangle_{\beta}^c \;=\;- \frac{\varepsilon^3}{g_s^2}\;,
\label{confD5}
\ee
matches its left-hand side, as shown in Eq.~(\ref{lefty}). Similarly, one could also discuss the conformal dimension of $\;T^i_{~i} = -p\;$ and find agreement between both sides of Eq.~(\ref{confD2}).
Finally,  Eq.~(\ref{confD4})  makes clear the expected relationship between
the stress--energy tensor and the central charge~(\ref{center1}),
$\;\langle T^{0}_{~0}(\vec{x}) T^{0}_{~0}(0)\rangle_{\beta}^c \sim 1/g_s^2\sim c\;$.

\section{Correspondence to an asymptotically de Sitter Universe}

We will now set up the  correspondence  between dS space and the theory of the thermal scalar in a similar manner to that of AdS/CFT  \cite{GKP,WittenAdS}, but yet with significant differences. To establish our proposed correspondence, it will be shown that  the HH wavefunction $\Psi_{HH}$  of an asymptotically dS Universe  can be calculated using the partition function of the CFT of the thermal-scalar condensate. The same CFT can be viewed as ``living'' on a spacelike surface which should also be regarded  as the future boundary of its  asymptotically dS dual.

Here, we are considering a situation in which an asymptotically dS spacetime decays into a radiation-dominated Universe. From the perspective of the microscopic string state, this corresponds to the phase transition from the Hagedorn phase of long strings to a thermal state of radiation. As argued in \cite{dSHigh}, we do expect the Hagedorn phase to be unstable, due to either a process which is similar to Hawking radiation or else to some coherent perturbation. From the viewpoint of the semiclassical spacetime, this decay corresponds to  the reheating of the Universe after inflation. The correlation functions then become temperature perturbations and are the late-time observables, just as in the standard inflationary paradigm. Meaning that the  late-time,   Friedmann--Robertson--Walker (FRW) observers  are the ``metaobservers'' \cite{dsCFTWitten}
or ``score-keeping observers'' \cite{dSHigh}
of the early inflationary epoch.

\begin{figure}[bth]
  \centering
  \vspace{-0.45in}
  \hspace{-.55in}
  \includegraphics[height=.45\textheight]{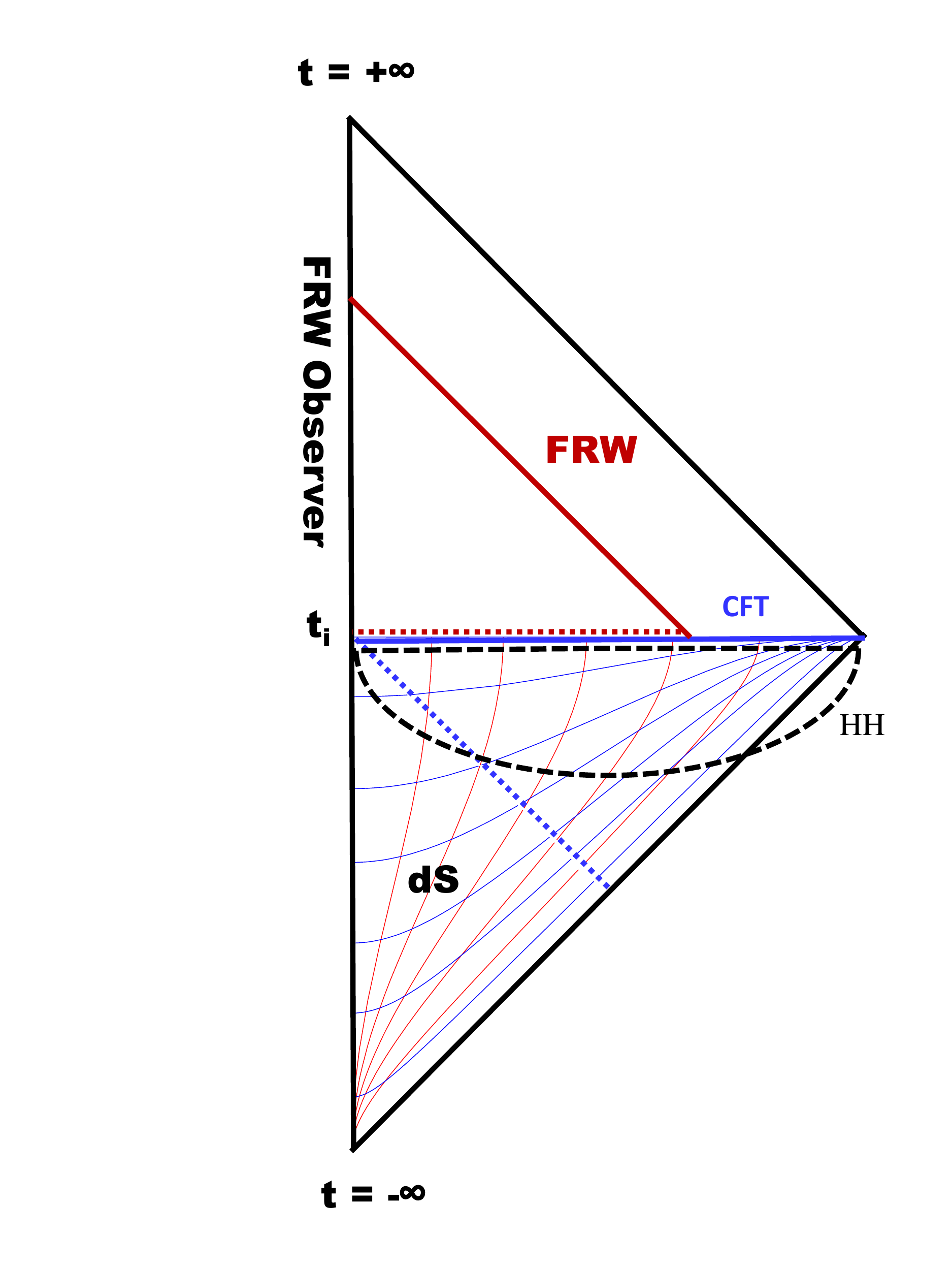}
  \caption{The correspondence between the CFT and dS space. The HH wave function is calculated on a Euclidean section of a $d+1$-dimensional space, as depicted by the black, dashed semicircle, while the Euclidean CFT is $d$-dimensional and ``lives" on the future boundary of dS, as depicted by the solid, blue line. In the upper half, the late observer's past light cone is displayed by the solid, red line, while in the lower half, lines of constant planar-dS coordinates $t$
and $r$ are shown in red (approximately vertical)  and blue (approximately horizontal), respectively.
  }\label{Fig:penrose1}
\end{figure}

As the FRW evolution starts in a thermal state,  an FRW observer might be compelled to invent a  prehistory to explain  the observable Universe. This is similar to  the way that a semiclassical observer invents a description of the BH interior \cite{BHfollies} (and see below). Three possible such prehistories are shown in Fig.~\ref{Fig:penrose1}.

An FRW observer would then  conclude that the Universe exponentially expanded during some epoch in its pre-history, for which  the inflationary paradigm provides a possible explanation. But let us emphasize the essential  point that the inflationary paradigm is an invented effective history of the Universe. What is physically real are the results of the measurements that are made by an FRW observer after the end of inflation \cite{dSHigh}.

It is interesting to compare the just-discussed cosmological picture to the corresponding situation in the case of BHs. In the  latter case, it is clear that an asymptotic, external observer is the one  who can eventually measure observables using the quantum state of the emitted radiation and is, therefore,
the score keeper for the interior.  The cosmological analogue --- perhaps not quite as obvious ---   is the late-time or FRW  observer.  The distant past of this observer, before the beginning of the hot-radiation phase, is the analogue of the BH interior. We similarly  argued for the case of BHs \cite{BHfollies} (also see \cite{nima})  that all   proposals for the pre-history are perfectly acceptable as long as they are  self-consistent, able to reproduce the  observable Universe and compatible with the laws of physics. By this line of reasoning, the puzzles of the FRW observer originate from trying to explain what is an intrisically quantum initial state in terms of effective semiclassical physics. The same situation was prevalent for BHs  and led to the  infamous BH paradoxes. As will be shown here, the FRW observer can interpret what is a maximally entropic  state as one of vanishing entropy  with an approximate description in terms of  the flat-space slicing of a classical  dS spacetime.

Let us briefly review the original proposal, first put forward by Witten \cite{dsCFTWitten}  and later by  Maldacena \cite{maldacenadS} (also see \cite{HarloStanford}), that the equality between the HH wavefunction of an asymptotically dS Universe and the partition function of some CFT should serve as a requirement for setting up a dS/CFT correspondence.   The idea was to start with  a Euclidean AdS spacetime but regard the direction perpendicular to the boundary --- which  is the radial coordinate in  AdS space ---  as the  time coordinate  in a Euclidean dS spacetime.
However, to the best of our knowledge, this idea was never explicitly realized in a way that is directly related to, or consistent with string theory. \cite{nogo}.
The suggested equality
$\;
\Psi_{HH}(g_{ij},J) = {Z}_{CFT}(g_{ij},J)\;
\label{HH-FT}
$
relied on certain identifications: The $d$-dimensional metric   $g_{ij}$ represents, on the left, the reduction of the  $(d+1)$-dimensional dS metric on the spacelike boundary and,  on the right, the  metric of the CFT. As for $J$, its dS meaning is the  boundary values of fields (like the graviton)  which can be used to set initial conditions for their post-inflationary evolution, whereas its CFT meaning is the sources for the fields in the  CFT Lagrangian.

Correlation functions of operators in the CFT were supposed to be calculated in the standard way; as derivatives of the partition function with respect to the sources.  Given the above interpretation, these correspond on the dS side to the boundary values  of bulk expectation values of spacetime fields.
For example, if a dS scalar field $\phi$ is considered, then
$\;
\langle \phi^2 \rangle = \int \left[D \phi \right] \phi^2 \left|\Psi_{HH}(\phi)\right|^2\;,
$
whereas
$\;
\langle \phi^2 \rangle  =  \frac{\delta Z_{CFT}}{\delta J_\phi \delta J_\phi}_{|J_\phi=0} \;.
$

We will follow \cite{dsCFTWitten,maldacenadS} in taking  the bulk spacetime as being the Poincar\'e patch of dS space in planar coordinates and the ground state of the bulk fields as being in the Bunch--Davies vacuum. However, the identifications between
dS and CFT quantities  will be different. We will start by identifying the physical components of the two different stress--energy tensors, that of the asymptotically dS bulk and that of the CFT.  The perturbed Einstein equations in the bulk will then be used to find a relationship between dS metric perturbations  and  perturbations of  the CFT stress--energy tensor.  We cannot use the CFT metric for this purpose because it is a fiducial, unphysical field.
As for the stress--energy tensor of the CFT, it  cannot be obtained as the derivative of the Lagrangian with respect to such a fiducial metric. Rather, it has to be defined in terms of the energy density and the pressure of the strings.

Our current interest  is in the case of pure gravity, so that  the only relevant bulk fields are  the tensor and scalar  perturbations of the metric.   In what follows,  we will make the abstract equality $\;\Psi_{HH}=Z_{CFT}\;$ explicit and then use it to calculate  correlation functions of the relevant fields. The correlation functions are our ultimate interest because  these are what correspond to  observable physical quantities. We will compare our results to those of
 the standard inflationary paradigm \cite{mukhanov} and to those which use the HH wavefunction \cite{Halliwell:1984eu,Hartle:2010vi,Hertog:2013mra}.

\subsection{Parameters and fields}

We now proceed  by comparing the dimensional parameters and dynamical fields  of  the thermal-scalar CFT with those of an asymptotically dS spacetime. As listed in Table~\ref{tparam}, each side contains a pair of  dimensional parameters:
The $D$-dimensional Newton's  constant $G_D$ and  the Hubble parameter $H$ in dS space
versus  $g_s^2$ and $\varepsilon$ on the CFT side.
 It should be noted that
 the string length scale $l_s$, or equivalently, the inverse of the  Hagedorn temperature, is a unit length rather than a dimensional parameter and
the temperature $T$  is not an additional parameter because it can be expressed in terms of $\varepsilon$ and  $T_{Hag}$, $\;T= \varepsilon+T_{Hag}\;$.
\begin{table}[!h]
  \centering
  \begin{tabular}{c|c}
  dS & FT  \\
  \hline
   $G_D$& $ g_s^2$\\
 $H$ & $\varepsilon$ \\
\end{tabular}
  \caption{Dimensional parameters in  dS space and the thermal CFT.}
  \label{tparam}
\end{table}

In the case of a pure theory of gravity in the asymptotically dS bulk, each side also contains two dynamical fields.   For dS space, these are the transverse--traceless (TT) graviton $h_{\mu\nu}$ and the scalar perturbation $\zeta$. Strictly speaking, $\zeta$ is dynamical only when the dS symmetries are broken, as it would be for a non-eternal asymptotically dS spacetime. For the CFT, the dynamical fields cannot simply  be the corresponding metric  perturbations, as  already discussed. Hence, we will consider  TT  and  suitably defined scalar perturbations of the CFT  stress--energy tensor and then, with the help of Einstein's equations, use these to deduce the corresponding perturbations of the dS metric. Table~\ref{tfields} includes  the corresponding  pairs of dynamical fields along with each pair's respective cosmological observable. There and subsequently, we have  denoted generic tensor perturbations of the CFT stress--energy tensor by $\delta \rho_{ij}$ and their TT components by $\delta\rho_{ij}^{TT}$.
\begin{table}[!h]
\centering
\begin{tabular}{c|c|c|}
dS  & CFT  & CO \\ \hline
   $h_{ij}$ &   $\delta \rho^{TT}_{ij}$ &   $P_T$ \\
$\tfrac{1}{H}\tfrac{\partial\zeta}{\partial t} $&  $\tfrac{\delta s}{s} $   & $P_\zeta$ \\
\end{tabular}
\caption{Fields and cosmological observables (CO). The quantity $\delta\rho^{TT}_{ij}$ is defined below in the text.}
\label{tfields}
\end{table}

In our framework, the  dynamical CFT  fields are  given in terms of  either the entropy perturbations $\delta s$ or the closely related perturbations of the energy density and pressure, $\;\delta\rho=\delta p=\varepsilon\delta s\;$, with the equalities following from the equation of state and first law respectively. Local scalar  perturbations in the entropy, energy and pressure  are not invariant under conformal transformations (rescalings  in particular)  and therefore do not constitute physical observables. The identity of the physical scalar perturbations will be clarified  in Subsection~4.3.2.  Similarly, vector perturbations are not physical, as these  can be undone by special conformal transformations. On the other hand,  TT tensor perturbations are physical.   Higher-spin perturbations ---  such as sextupole, hexapole, {\em etc}. ---  will involve derivatives  as these  are the only other vectors available in the CFT. So that, for length scales larger than the horizon,
$\;k\ll H\;$,  such higher-order perturbations are suppressed.

As for the TT components of the perturbations
of the stress--energy tensor,  on the basis of isotropy,  each independent mode
 fluctuates with equal strength and  the sum of their squares is equal to the  square of the energy-density perturbation,
$
\sum\limits_{i,j} \;|\delta\rho^{TT}_{ij}|^2=\tfrac{1}{2} (d+1)(d-2)|\delta\rho^{TT}_{ij}| =|\delta \rho|^2\;
$.
For sake of completeness, the TT components can  be  formally defined in terms of a  transverse projection operator $P^T_{lm}$,
\be
P^T_{lm}\;=\;\left(\delta_{lm}-\tfrac{\nabla_l\nabla_m}{\nabla^2}\right)\;,
\label{einsteinz}
\ee
which leads to the construction of a TT projector in the standard way,
\be
\delta\rho^{TT}_{ij}\;=\; \left(P^T_{il}P^T_{jm}-\tfrac{1}{d-1}P^T_{ij}P^T_{lm} \right)\delta\rho^{lm}\;.
\ee

Using the above correspondence between the two sets of fields and dimensional parameters, we can turn the relationship between the HH wavefunction and the CFT  partition function
into a more explicit equality,
\be
\Psi_{HH}\left(h_{ij},\zeta ;G_D,H\right) \;=\; {\cal Z}_{CFT}\left(\delta\rho^{TT}_{ij}, \frac{\delta s}{s} ; {g}_s^2, \varepsilon\right)\;.
\label{HH-FT-1}
\ee

\subsection{Thermodynamics}

The objective here is to make the correspondence between the CFT and dS space more precise  by comparing  their respective  values for  the entropy.
As for other possible comparisons, the Gibbons--Hawking value of the dS temperature  $\;T_{dS}= \frac{H}{2\pi}\;$, is not directly related to observables in the FRW epoch because of its  observer dependence. The energy density is indeed observable  but even more ambiguous, as the original derivation of the Gibbons--Hawking  entropy was for a closed dS space for which  the total energy vanishes  \cite{GHdS}.   Our expectation is that the energy density of the strings will increase as the Hagedorn transition proceeds, until it becomes comparable to the  Hagedorn energy density. Hence, it is the entropy that serves as the  most reliable observable for comparison purposes.

Let us now  recall from Eq.~(\ref{consol}) that  the CFT entropy density
is given by
$
s_{CFT}\;=\;  \frac{\varepsilon}{g_s^2}\;,
$
while also recalling that $\varepsilon$ is the associated (effective)
temperature as in Subsection~3.4. The entropy of the CFT in a Hubble volume $V_d(H)$ (or ``causal patch") is then
\be
S_{CFT}\;=\;\frac{\varepsilon V_d(H)}{g_s^2}\;,
\label{scft}
\ee
which  should be compared to the Gibbons--Hawking entropy on the
dS side \cite{GHdS},
\be
S_{dS}\;=\;\frac{A_d(H)}{4 G_D}\;=\; \frac{H V_d(H)}{4G}
\;,
\label{sghds}
\ee
where $A_d(H)$ is the surface area of the Hubble volume and
$\; A_d(H)= HV_d(H)  \;$ in planar coordinates has been used.

Equating the two entropies,
\be
S_{CFT}\;=\;S_{dS}\;,
\label{scftdS}
\ee
we then obtain
\bea
\label{gsH}
\frac{8\pi G_D}{g_s^2} \frac{\varepsilon}{H} \;=\; 2 \pi\;.
\eea
Recall that  we have absorbed  numerical, string-theory dependent, factors into $\varepsilon$ and $g_s^2$ (see Section~2). Making these factors explicit, one could then fix the ratio $\frac{8\pi G_D}{g_s^2}$ in any specific
string theory, which would in turn  fix the  ratio $\frac{\varepsilon}{H}$. However, as the relation between $G_D$ and $g_s^2$ is highly model dependent,
 a detailed discussion on these ratios will be deferred  to a
future investigation.

Given the identity in Eq.~(\ref{scftdS}), the expected relation \cite{GHdS}
\be
\left|\Psi_{HH}\right|^2 \;=\;  e^{+S_{dS}}\;
\ee
can now be recovered from the equilibrium value of the  CFT partition function
\be
 Z^2_{CFT}\;=\; e^{-2\frac{1}{T}\mathcal{S}_0}\; =\; e^{+\frac{2}{T} \int d^d x\; \frac{1}{2}\frac{\varepsilon^2}{g_s^2}}\;,
\ee
where the right-most exponent follows from Eq.~(\ref{sos}) and the use of flat, planar coordinates. One should take note of
the crucial sign change of the exponent thanks to the negativity of $\mathcal{S}_0$.
For the purposes of matching this partition function to the HH wavefunction, we need to change  the  prefactor
in the exponent  from $1/T$  to $1/\varepsilon$. This is consistent with the perspective of Subsection~3.4 and is, once again, related to the effective temperature of the long strings being equal to $\varepsilon$ rather than
the  microscopic temperature of the strings $\;T\sim T_{Hag}\;$. The end result is
\bea
\left|\Psi_{HH}\right|^2\;= \;Z_{CFT}^2(T\to \varepsilon)
&=& \exp\left(\frac{1}{\varepsilon}\int d^d x~\frac{\varepsilon^2}{g_s^2} \right) \cr
&=& \exp\left(\int d^d x~ s\right) \cr &=& \exp\left(S_{CFT}\right)\cr &=& \exp\left(S_{dS}\right)\;,
\eea
where the integral is over the Hubble volume and Eq.~(\ref{scftdS}) has been used at the end.

It should be  emphasized that, in spite of the exponentially growing magnitude  of the wavefunction, the perturbations are well behaved and controlled by a
well-defined Gaussian integral as in Eqs.~(\ref{gauss1}) and~(\ref{gauss2}).

Our definition of the HH wavefunction in terms of $Z_{CFT}$ resolves several longstanding issues about this wavefunction and its use in Euclidean quantum gravity \cite{EQG1,EQG2}. Formally, the  Euclidean gravitational  action is unbounded from below, and the integral defining it is badly divergent. But the wavefunction  is certainly relevant to perturbations about an asymptotically dS space and, as we have seen, the  associated Gaussian integral  is itself  well defined and convergent. Moreover, from our perspective, the growing exponential for the magnitude of the wavefunction is not a vice but a virtue, as it is  needed to explain the large entropy of dS space. Additionally, if  $\Psi_{HH}$ is viewed  as defining a probability distribution for a background dS Universe, the distribution is peaked at  small values of the cosmological constant, thus implying  a large and empty universe which  disfavors inflation. Our definition of the wavefunction, on the contrary, predicts a large, hot Universe in lieu of inflation. Finally, our definition extends the domain of the quantum state of the Universe beyond the semiclassical regime and demonstrates that the resolution of the initial singularity problem must rely on strong quantum effects.

\subsection{Two-point correlation functions and spectrum of perturbations}

We begin this part of the analysis  by  expanding the Lagrangian ${\cal L}_E (s)$ in Eq.~(\ref{lag1}) about the equilibrium solution $s_0$ up to second order in  the perturbation
strength $\;\delta s(\vec{x})= s(\vec{x}) - s_0$\;.  This will enable us to   calculate the  two-point correlation functions of the CFT,  which can be used in turn to calculate the  spectra of  the corresponding cosmological observables.

The relevant term in the just-described expansion is the quadratic term,
\be
\mathcal{S}_E^{\;(2)}\;=\; \frac{1}{T} \int d^d x \dfrac{1}{2} g_s^2 \delta s^2 +\cdots\;,
\ee
from which it follows that
\be
\langle \delta s(\vec{x}) \delta s (0) \rangle \;=\;\int [{\cal D}\delta s]\; \delta s(\vec{x}) \delta s (0)\;  e^{-\frac{1}{T} \int d^d x \frac{1}{2} g_s^2 \delta s^2}\;=\;\frac{T}{g_s^2} \delta^d(\vec{x})\;.
\label{deltas2}
\ee

In cosmology, it is customary to use the power spectrum of the two-point function as the observable quantity. What is then required is the Fourier transform  of the perturbation $\delta s_{\vec{k}}$,
which is  related to $\delta s(\vec{x})$ in the usual  way,
\be
\delta s(\vec{x})\;=\;\frac{1}{(2\pi)^d} \int d^d k\; e^{i\vec{k}\cdot \vec{x}}\; \delta s_{\vec{k}}\;.
\label{fauria}
\ee
The two-point function for  $\delta s_{\vec{k}}$ is expressible as
\be
\langle \delta s_{\vec{k}_1} \delta s_{\vec{k}_2} \rangle\; =\;|\delta s_{\vec{k}_1}|^2 (2\pi)^d \delta^d(\vec{k}_1+\vec{k}_2)\;,
\label{ss1}
\ee
where
\be
|\delta s_{\vec{k}}|^2 \;=\; \frac{T}{g_s^2}\;
\label{ss2}
\ee
can be deduced from Eq.~(\ref{deltas2}).

Now applying the standard relationship between a power spectrum and its associated two-point function,
\be
d(\ln{k})\; P_{\delta s}(k)\;=\; \frac{d^d k}{(2\pi)^d} |\delta s_{k}|^2\;,
\ee
we  obtain the spectral form
\be
P_{\delta s} (k)\; =\; \frac{d\Omega_{d-1} k^d}{(2\pi)^d}\; \frac{T}{g_s^2}\;,
\label{pdeltas}
\ee
where $d\Omega_{d-1}$ is the solid angle subtended by a $(d-1)$-dimensional spherical surface.
The power spectrum  has, by definition,  the same dimensionality as
$\langle\delta s(\vec{x})^2\rangle$, and this fixes the power of $k$ unambiguously.

Since $\;\delta \rho = \varepsilon\delta s\;$ from the first law and
 $\;\delta p=\delta \rho\;$  from the equation of state,
it can also be deduced that
\be
P_{\delta \rho} (k) \;=\;
P_{\delta p} (k)\;=\;
\frac{d\Omega_{d-1}k^{d}}{(2\pi)^d} \frac{T\varepsilon^2}{g_s^2}\;.
\label{pdeltarho}
\ee

\subsubsection{Tensor perturbations}

To obtain the power spectrum of the tensor perturbations,
we start with the relationship between a specific polarization of the tensor perturbations of the metric and the corresponding component of the stress--energy tensor perturbation  (see, {\em e.g.}, \cite{mukhanov}),
\bea
\langle|h_{ij}(k)|^2\rangle_{dS} &=& \frac{(4\pi G_D)^2}{(k^2)^2} \langle|\delta T^{TT}_{ij}(k)|\rangle^{2}_{\;dS}\; \nonumber \\
&=&
\frac{(4\pi G_D)^2}{(k^2)^2} \langle|\delta \rho^{TT}_{ij}(k)|\rangle^{2}_{\;CFT}\;,
\label{einsteiny}
\eea
where the proposed duality has been applied in the second line  and
thus the validity of the second equality
only applies on the spacelike matching surface ({\em i.e.}, on  the
future boundary of the asymptotically dS spacetime).

Let us recall that
$\;\sum\;\langle|\delta \rho^{TT}_{ij}(k)|\rangle^{2}_{\;CFT}=|\delta\rho|^2_{\;CFT}\;$.
Then, from Eq.~(\ref{einsteiny}), it follows that $\sum\;\langle|h_{ij}(k)|^2\rangle_{dS}$ can be directly related to $|\delta\rho|^2_{\;CFT}\;$, and  one can
similarly  relate the  total power spectrum for the tensor perturbations $P_T(k)$
to
the spectrum in Eq.~(\ref{pdeltarho}),
\bea
P_T(k)_{|k\to H,T\to\varepsilon} &=& \frac{(4\pi G_D)^2}{ (k^2)^2}  P_{\delta\rho}{}_{|k\to H,T\to\varepsilon} \nonumber \\
&=& \tfrac{1}{4} (8\pi G_D)^2  \frac{\varepsilon^3}{g_s^2}  H^{d-4} \frac{d\Omega_{d-1}}{(2\pi)^d}\;,
\eea
where the standard horizon-crossing condition $\;k\to H\;$ has been applied  and our  usual replacement  $\;T\to \varepsilon\;$ has been made.

Next, using Eq.~({\ref{gsH}}), we obtain
\be
P_T(H)\;=\;\frac{\pi}{2} \frac{\varepsilon^2}{H^2} (8 \pi G_D) H^{d-1} \frac{d\Omega_{d-1}}{(2\pi)^d}\; \;,
\ee
or,  in terms of the dS entropy in Eq.~(\ref{sghds}),
\be
P_T(H)\;\sim\;  \frac{1}{S_{dS}} \;,
\ee
as expected. Notice that $P_T(H)$ is dimensionless.

In the observationally relevant case of $\;d=3\;$, the above reduces to
\be
P_T(H)\;=\; \frac{1}{4\pi} \frac{\varepsilon^2}{H^2}  \frac{H^{2}}{m_P^2}\;,
\ee
which, has the same parametric dependence as the standard inflationary result,
\be
P_T({\rm inflation})\;=\; \frac{2}{\pi^2}  \frac{H^{2}}{m_P^2} \;.
\ee
A calculation of the tensor power spectrum using the HH wavefunction with an additional scalar field \cite{Halliwell:1984eu,Hertog:2013mra} is in  agreement with the standard inflationary outcome  and, therefore, our result is  also in qualitative agreement with this calculation.

It should be emphasized that  we assumed in the calculation that the state is
one of  exact thermal equilibrium, so that its temperature is uniform or,
equivalently, $\;\varepsilon(k)= constant$. It is for this reason that the
spectrum of tensor perturbations was found to be exactly scale invariant. It may well be that the effective temperature of the state is not exactly constant and could be scale dependent due to some source of conformal-symmetry breaking.
This breaking is quite natural insofar as  the state has a finite extent; equivalently, the dS spacetime is non-eternal. Nevertheless, the breaking is expected to be quite small, as its effects are  proportional to the deviations of the spacetime from an eternal dS background. We will discuss this issue further after discussing the scalar perturbations.

\subsubsection{Scalar perturbations}

In an eternal asymptotically dS space,  time does not exist and it is impossible for a single observer to see the extent of the whole state. By contrast,
in a non-eternal asymptotically dS spacetime, a quantity that measures time ---  a ``clock'' ---  can be introduced. The same  must apply to each of their respective CFT duals.  For instance, in semiclassical inflation, the clock  is introduced in the guise of a slowly rolling inflaton field. On either side of
our proposed correspondence, the clock is the total {\em observable} entropy of the state in units of the horizon entropy. And it
is the fluctuations in this clock time that serves as the  dual to the scalar
modes  of dS space, as we now explain.

To formulate the dual of the gauge-invariant scalar perturbations $\zeta$ \cite{mukhanov}, we will follow \cite{dSHigh} and rely on the relationship between $\zeta$ and the perturbations in the number of e-folds $\delta N_{e-folds}$. This method was previously used to calculate super-horizon perturbations in the ``separate Universe" approach and  the $\delta N$ formalism  \cite{separate1,separate2}, where it was shown that
\be
\zeta \;=\; \delta N_{e-folds}\;.
\label{zeta}
\ee
It should be emphasized that Eq.~(\ref{zeta}) fixes completely the normalization of $\zeta$. From our perspective, what is important is that   the value of $\delta N_{e-folds}$ can be expressed in terms of CFT quantities, as we will clarify in the ensuing discussion.

The number of e-folds that an FRW observer has to postulate is, from his perspective,  determined by the increase in volume which  is required to explain the difference in  entropy between that in a single Hubble horizon $\;S_{H}\sim S_{dS}\;$ and the total  entropy of the Universe $\;S_{tot}=n_HS_H\;$. From this observer's perspective, the parameter $n_H$ is the number of causally disconnected Hubble volumes $V_H$ at the time of reheating; that is,
\be
\label{boblawblaw}
n_H \;=\; e^{d \;N_{e-folds}}\;=\; \frac{V_{tot}}{V_{H}}\;=\;\frac{S_{tot}}{S_H}\;,
\ee
where the last equality assumes that there are no additional entropy-generating mechanisms  after the inflationary period (otherwise, the final ratio would be an upper bound) and that $S_H$ is constant, independent of its location.
Meanwhile, a  hypothetical CFT observer  faces the analogous task of accounting for an extremely large total entropy after the phase transition from strings to radiation.

To make use of the relationship between $\delta N_{e-folds}$ and $\zeta$, we
call upon a known  expression for $\zeta$ in terms of  pressure perturbations \cite{separate2},
\be
\frac{1}{H} \frac{\partial \zeta}{\partial t} \;=\; -\frac{1}{p+\rho} \delta p_{|\rho}\;.
\ee
Then, since $\;p+\rho=\varepsilon s\;$ and $\;\delta p =\delta\rho= \varepsilon \delta s\;$,
\be
\frac{1}{H} \frac{\partial\zeta}{\partial t}\; =\; -\frac{\delta s}{s}\;.
\ee

Next, the conformal symmetries on either side of the duality allows for the replacement of  $\frac{1}{H} \frac{\partial}{\partial t}$ with
$-\frac{\partial}{\partial(\ln k)}$,
\be
\frac {\partial\zeta}{\partial(\ln k)}\; =\; \frac{\delta s}{s}\;,
\label{zeta1}
\ee
or, formally,
\be
\zeta \;=\; \int d(\ln k) \;\frac{\delta s}{s}\;.
\label{zeta2}
\ee
This result can be recast as
\be
\zeta \; = \;  \int \frac{d(\ln V)}{d} \;\frac{\delta s}{s}\; = \; \frac{1}{d} \int d^dx \;\frac{\delta s}{V s} \;=\;\delta N_{e-folds}\;,
\ee
where the first equality follows
from conformal symmetry  and the last one from Eq.~(\ref{boblawblaw}).

We can now call upon   Eq.~(\ref{zeta1}) for $\zeta$ and the equilibrium value
for $s$ in Eq.~(\ref{consol}) to show that the two-point function for the scalar perturbations satisfies
\be
\frac{\partial}{\partial(\ln k_1)} \frac{\partial}{\partial(\ln k_2)}
\langle  \zeta_{\vec{k}_1}  \zeta_{\vec{k}_2} \rangle\;=\;
\left(\frac{g_s^2}{\varepsilon}\right)^2 \;\langle \delta s_{\vec{k}_1} \delta s_{\vec{k}_2} \rangle\;.
\label{zeta-twpoint}
\ee
Observing that both sides of Eq.~(\ref{zeta-twpoint}) are of the  form $f(k_1) \delta^d(\vec{k}_1+\vec{k}_2)$,
one can integrate twice over both sides  and compare the coefficients. The result is
\be
\langle |\zeta_{k}|^2\rangle \;=\; \frac{N_{e-folds}}{d} \left(\frac{g_s^2}{\varepsilon}\right)^2\langle |\delta s_{k}|^2\rangle \;=\; \frac{N_{e-folds}}{d} \frac{T g_s^2}{\varepsilon^2}  \;,
\ee
where the second equality follows from Eq.~(\ref{ss2}) and
the factor of $N_{e-folds}$ results from one of the  integrals on the right, $\;-\int d(\ln k)=\int H dt =\int d(\ln a)= N_{e-folds}\;$.
The associated  power spectrum is then
\be
P_{\zeta}(k)\;=\;\frac{N_{e-folds}}{d} \frac{T g_s^2}{\varepsilon^2}\frac{k^d d\Omega_{d-1}}{(2\pi)^d}\;.
\ee

To make contact with the dS calculation, the conditions  $\;k\to H\;$ and $\;T\to \varepsilon\;$  can once again be imposed,
\be
 P_{\zeta}(H)\;=\; \frac{N_{e-folds}}{d} \frac{g_s^2}{\varepsilon} H^{d} \frac{d\Omega_{d-1}}{(2\pi)^d} \;.
\label{pzeta}
\ee
If we further substitute $8\pi G_D$ for $g_s^2$ using Eq.~(\ref{gsH}), then
\be
 P_{\zeta}(H)\;=\; \frac{N_{e-folds}}{2\pi d}\;  8 \pi G_D\; H^{d-1}\frac{d\Omega_{d-1}}{(2\pi)^d} \;.
\label{pzeta1}
\ee

The fact that $P_{\zeta}$ is enhanced by the number of e-folds with respect to the tensor perturbations is a significant feature of the correspondence,
\be
P_{\zeta}\;\sim\; N_{e-folds}P_T\;.
\ee
The enhancement factor  of $N_{e-folds}$  can be traced to the large size of the initial string state rather than to the scaling properties of the CFT or to deviations from scale invariance. This is unlike in models of semiclassical inflation, for which the tensor perturbations are viewed as suppressed with respect to their scalar counterparts by a factor that is explicitly related to the amount of deviation from scale invariance.

For the  $\;d=3\;$ case with $\;m_P^2=1/(8 \pi G)\;$,
\be
 P_{\zeta}(H)\;=\; \frac{N_{e-folds}}{4\pi^3 d} \frac{H^2}{m_P^2}\;,
\ee
which
can be compared to the standard inflationary result,
\be
P_{\zeta}(H)_{inflation}\;=\;\frac{1}{\epsilon_{inf}}\frac{1}{8\pi^2} \frac{H^2}{m_P^2}\;,
\ee
where  $\epsilon_{inf}$ parametrizes the deviation from scale invariance, $\;1-n_S= 6 \epsilon_{inf}-2\eta_{inf}\;$. Here, $n_S$ is the scalar spectral index
and $\epsilon_{inf}$, $\eta_{inf}$ are the slow-roll parameters. In simple models of inflation, $\;\epsilon_{inf}\sim 1/N_{e-folds}\;$; meaning that  our result is in qualitative agreement with that of semiclassical inflation.

A calculation of the scalar perturbations using the HH wavefunction \cite{Halliwell:1984eu,Hartle:2010vi,Hertog:2013mra} is in agreement with the standard inflationary result and, just like for models of  inflation, requires an additional scalar field to render the scalar perturbations as  physical.  Meaning that our result for the scalar power spectrum  is  in  qualitative agreement with the HH calculation as well.

An important observable is the tensor-to-scalar power ratio $r$. In general,
\be
r \;=\;\frac{P_T}{P_\zeta}\;=\;\frac{d}{N_{e-folds}}\frac{\pi^2 \varepsilon^2}{H^2}\,
\label{r-eq}
\ee
and, in the $\;d=3\;$ case,
\be
r \;=\; \frac{3}{N_{e-folds}}\frac{\pi^2 \varepsilon^2}{H^2}\;.
\ee
Given that  $\;\varepsilon\sim H\;$ as expected, the above value of $\;r\sim 1/N_{e-folds}\;$ would
correspond to a high scale of inflation  if interpreted within simple
models of semiclassical inflation. This is consistent with our expectation that the energy density is of the order of $T_{Hag}^4$ \cite{dSHigh}.

\subsection{Higher-order correlation functions and deviations from scale-invariance}

The discussion has, so far, been focusing on the  quantities that are the least sensitive to the choice
of model; namely, the two-point functions in the case of  conformal invariance.
 Our results could be extended to more model-dependent quantities, such as two-point correlation functions when conformal invariance is weakly broken or higher-point functions for the conformally invariant case. We will not extend the calculations at the
present time but do  anticipate a more detailed analysis
along this line  in the future.  Let us, meanwhile, briefly explain the significance of such  model-dependent calculations.

Deviations from conformal invariance can arise from spatial dependence (equivalently, $k$ dependence) of the effective temperature $\varepsilon$ or  the
string coupling  $g_s^2$ or both. These will in turn introduce scale dependence into the tensor and scalar power spectra. The scale dependence is an observable feature; however, because
of its dependence on the details  of the background solution and on the nature of the Hagedorn transition --- and not just on scales and symmetries --- it is,  in some sense, a less fundamental aspect of the correspondence.

The higher-order terms in the  CFT Lagrangian, as  discussed in  Eqs.~(\ref{SHO34}-\ref{SHOgfin}),  are also  present  when the conformal symmetry remains unbroken. However, these terms are still model dependent  as they depend on the specific string theory. But, in spite of their relative smallness, they remain  of considerable interest, as such terms  can be used to calculate three-point (and higher) correlation functions. These multi-point correlators are what determines the non-Gaussianity of the spectra of perturbations and, therefore, represent an opportunity for distinguishing our proposed correspondence from the standard inflationary paradigm.  Unfortunately, it is already quite evident that such effects are small.

\section{Conclusion and outlook}

We have put forward a new correspondence between asymptotically dS space and a CFT dual by showing that the partition function of the CFT is equal to the HH wavefunction of the dS space. Our correspondence provides a complete qualitative description of a non-singular initial state of the Universe and, in this sense, replaces the big-bang singularity and semiclassical inflation.

We have built off of a previous work \cite{dSHigh} which shows that an asymptotically  dS spacetime has a dual description in terms of a state of interacting, long, closed, fundamental  strings in their high-temperature Hagedorn phase. A significant, new development was the identification of the entropy density of the strings with the magnitude-squared of a condensate of a thermal scalar whose path integral is equal, under certain conditions, to the full  partition function for the Hagedorn phase of string theory.  The strings are thus described by a  thermal CFT, which can also  be viewed as a Euclidean field theory that has been compactified on a string-length thermal circle.  Surprisingly,  the  reduced theory  has the scaling properties of a two-dimensional CFT in spite of formally being defined in a manifold with $\;d\geq 3\;$ spatial dimensions.

Our correspondence  provides a clear origin for the entropy of dS space as
the microscopic entropy of a hot state of strings. This explanation clarifies how a state whose equation of state is $\;p=-\rho\;$, as in dS space, can have any entropy at all when  the thermodynamic relation $\;p+\rho=sT\;$ suggests that both the entropy and the temperature are vanishing.  From the stringy point of view, the pressure is rather maximally positive and the negative  pressure of dS space is an artifact of insisting on a semiclassical geometry when none is justified.

The proposed duality redefines the HH wavefunction and resolves several outstanding issues with its common interpretation, such as the divergence of the Euclidean path integral and its preference for an empty Universe with a very small cosmological constant.

We have shown how the  power spectra for the  tensor and scalar  perturbations
of the asymptotically dS metric can be calculated on the CFT side of the correspondence  by identifying
the two dual fields,  the scalar and tensor perturbations
of the CFT  stress--energy tensor. As was  discussed in detail, these calculations  reproduce,  qualitatively, the results of the standard inflationary paradigm and the corresponding calculations which use the HH wavefunction. Although any specific set of predictions will depend on the value of  an order-unity number ---
the ratio of  the effective temperature of the string state $\varepsilon$ to the Hubble parameter $H$ ---
our framework does   provide an opportunity to compare the predictions of specific string-theory-based  models for cosmological observables. In addition, the
strength of the  scalar perturbations was found to be naturally enhanced by a factor of $N_{e-folds}$,
even when the  theory formally exhibits local scale invariance.  This places our predicted
 value for the cosmological observable $r$ well within the empirical bounds.

Let us  now finish by discussing  some remaining issues  and possible extensions of the current analysis:

First, it should be reemphasized that we do not explain why the Universe is large. The  entropy of the string state is large because this corresponds to a large asymptotically dS Universe and thus leads to  a large FRW Universe in the state's future. The value of the entropy should be viewed as  part of the definition of the initial state.

Still lacking is a qualitative description of just how the state of hot strings decays into the  state of hot radiation which follows; a transition which is known as
reheating in inflation. In our case, the transition corresponds to a phase transition between the Hagedorn phase of long strings and a phase of short strings propagating in a semiclassical background. Because of the close parallels between early-Universe cosmology and BHs,
our expectation is that the transition is described by a decay mechanism that is
akin to Hawking radiation.

Our proposal can be extended to incorporate the effects of deviations away from conformality.  To make such a calculation precise, the issue of how the effective temperature $\varepsilon$ and the coupling $g_s^2$ depend on scale will have to be resolved. It will also, as mentioned,  be necessary to fix the ratio $\varepsilon/H$, which amounts to understanding the exact relation between the string coupling and Newton's constant in specific compactifications of various string theories.  Another possible extension is the incorporation of three-point correlation functions and higher. This entails the inclusion of
 yet-to-be-specified higher-order terms in the CFT Lagrangian, and using these to calculate three- and higher-point correlation functions in dS space.
Yet another interesting extension is to include  other  dynamical fields besides the physical graviton modes and their CFT dual; for instance, the dilaton of the underlying string theory.

The connection between  our proposed correspondence and the  AdS/CFT correspondence is not currently clear. What is clear, though, is that if such a connection exists, it  must  differ from previous proposals which regard the AdS radial  direction as Euclidean time and AdS time as one of the spatial coordinates. It is still possible that the two frameworks are somehow connected at a mathematical level or even at a yet unknown deeper conceptual level.

\section*{Acknowledgments}

We would like to thank
Sunny Itzhaki, Volodya Kazakov, Kyriakos Papadodimas, Riccardo Rattazzi, Misha Shaposhnikov, Marko Simonovic,  Gabriele Veneziano, Sasha Zhiboedov  and, in particular, Toni Riotto and  Yoav Zigdon
for useful discussions and suggestions. We would also like to thank Thomas Hertog, Heliudson de Oliveira Bernardo and Kostas Skenderis for clarifying discussions on their work.  The research of RB   was supported by the Israel Science Foundation grant no. 1294/16. The research of AJMM received support from
an NRF Evaluation and Rating Grant 119411 and a Rhodes  Discretionary Grant RD38/2019.
RB thanks the TH division, CERN where part of this research was conducted. AJMM thanks Ben Gurion University for their  hospitality during his visit.

\end{document}